\documentclass[twocolumn, tighten]{aastex631}

\shorttitle{Comparing Quantiles}
\shortauthors{Thorp et al.}

\usepackage{amsmath}	
\usepackage{bm}
\usepackage{booktabs}
\usepackage[normalem]{ulem}

\DeclareMathOperator{\diag}{diag}

\DeclareMathOperator{\var}{var}

\defcitealias{alsing24}{A24}

\begin{document}

\title{Data-Space Validation of High-Dimensional Models by Comparing Sample Quantiles}

\correspondingauthor{Stephen Thorp}
\email{stephen.thorp@fysik.su.se}

\author[0009-0005-6323-0457]{Stephen Thorp}
\affiliation{The Oskar Klein Centre, Department of Physics, Stockholm University, AlbaNova University Centre, SE 106 91 Stockholm, Sweden}

\author[0000-0002-2519-584X]{Hiranya V.\ Peiris}
\affiliation{Institute of Astronomy and Kavli Institute for Cosmology, University of Cambridge, Madingley Road, Cambridge CB3 0HA, UK}
\affiliation{The Oskar Klein Centre, Department of Physics, Stockholm University, AlbaNova University Centre, SE 106 91 Stockholm, Sweden}

\author[0000-0002-0041-3783]{Daniel J.\ Mortlock}
\affiliation{Astrophysics Group, Imperial College London, Blackett Laboratory, Prince Consort Road, London, SW7 2AZ, UK}
\affiliation{Department of Mathematics, Imperial College London, London SW7 2AZ, UK}

\author[0000-0003-4618-3546]{Justin Alsing}
\affiliation{The Oskar Klein Centre, Department of Physics, Stockholm University, AlbaNova University Centre, SE 106 91 Stockholm, Sweden}

\author[0000-0002-3962-9274]{Boris Leistedt}
\affiliation{Astrophysics Group, Imperial College London, Blackett Laboratory, Prince Consort Road, London, SW7 2AZ, UK}

\author[0000-0003-1943-723X]{Sinan Deger}
\affiliation{Institute of Astronomy and Kavli Institute for Cosmology, University of Cambridge, Madingley Road, Cambridge CB3 0HA, UK}



\begin{abstract}
We present a simple method for assessing the predictive performance of high-dimensional models directly in data space when only samples are available. Our approach is to compare the quantiles of observables predicted by a model to those of the observables themselves. In cases where the dimensionality of the observables is large (e.g.\ multiband galaxy photometry), we advocate that the comparison is made after projection onto a set of principal axes to reduce the dimensionality. We demonstrate our method on a series of two-dimensional examples. We then apply it to results from a state-of-the-art generative model for galaxy photometry (\texttt{pop-cosmos}) that generates predictions of colors and magnitudes by forward simulating from a 16-dimensional distribution of physical parameters represented by a score-based diffusion model. We validate the predictive performance of this model directly in a space of nine broadband colors. Although motivated by this specific example, we expect that the techniques we present will be broadly useful for evaluating the performance of flexible, non-parametric population models of this kind, and other settings where two sets of samples are to be compared.
\end{abstract}

\keywords{Astrostatistics techniques (1886); Principal component analysis (1944); Bootstrap (1906); Redshift surveys (1378); Galaxy photometry (611)}


\section{Introduction}
Comparing two distributions given only samples drawn from each is a ubiquitous problem in many fields of science. It is typically far easier to simulate mock data to replicate the observations than it is to evaluate the associated likelihood.

This situation is especially common in astronomy, where source catalogs are subject to complicated noise and selection effects.  Moreover, astrophysical simulations typically do not have a source-by-source correspondence with the catalogs, and they must hence be assessed on their ability to reproduce population-level characteristics of the data. The scale and fidelity of current and future surveys (e.g.\ COSMOS, \citealp{scoville07, weaver22}; the Kilo-Degree Survey, \citealp{dejong13, kuijken19}; the Dark Energy Survey, \citealp{des05, abbott18, sevillanoarbe21}; {\it Euclid}, \citealp{laureijs11, mellier24}; the Vera C.\ Rubin Observatory's Legacy Survey of Space and Time, \citealp{ivezic19}) is such that traditional parameterized population models are unlikely to be adequate. This has led to the increasing use of generative machine learning (ML) pipelines (e.g.\ \citealp{li23,moser24,alsing24}); these models have the necessary flexibility but typically do not have tractable likelihoods.

There is hence a need for  model checking and validation schemes that rely solely on the comparison of two high-dimensional samples (i.e.\ mock and real data). This, however, presents significant challenges, as we do not have access to the arsenal of Bayesian model checking tools, such as posterior predictive tests \citep{rubin84, meng94, gelman96} and cross-validation \citep{vehtari17}, which have become increasingly common in astronomy \citep[e.g.][]{protassov02, park08, huppenkothen13, stein15, chevallard16, feeney19, abbott19, rogers21, mcgill23, welbanks23, challener23, nixon23}. Other popular metrics for high-dimensional model comparison include the Kullback--Leibler divergence \citep{kullback51}, and the related Jensen--Shannon divergence \citep{lin91}, which are similarly reliant on having a well-defined density function for model parameters or data (for examples in a cosmology context, see e.g.\ \citealp{verde13, leclercq15, leclercq16, charnock17, hee17, nicola17, nicola19, moews19, pinho21}). If this is not available in closed form (e.g.\ from a Gaussian approximation to the data distribution or posterior; see \citealp{amara14, seehars14, grandis16a, grandis16b, schuhmann16}), high-dimensional density estimation is required to compute the divergence.

In univariate cases there are many applicable non-parametric two-sample statistical tests that do not rely on having a closed-form probability density: the Kolmogorov--Smirnov (K--S) test \citep{kolmogorov33, smirnov48}; the Cram\'er--von Mises test \citep{cramer28, vonmises28}; the Anderson--Darling test \citep{anderson52, darling57, pettitt76}; and quantile--quantile (Q--Q) and probability--probability (P--P) plots \citep{wilk68}. Of these, Q--Q and P--P plots have a number of desirable characteristics, as highlighted recently by \citet{eadie23} and \citet{buchner23}. A particular advantage is that they discard less information than a single numerical test statistic, and the two plots complement each other due to their sensitivity to different aspects of the samples (for a review see \citealp{fisher83}).

\begin{figure*}[ht!]
    \centering
    \includegraphics[width=\linewidth]{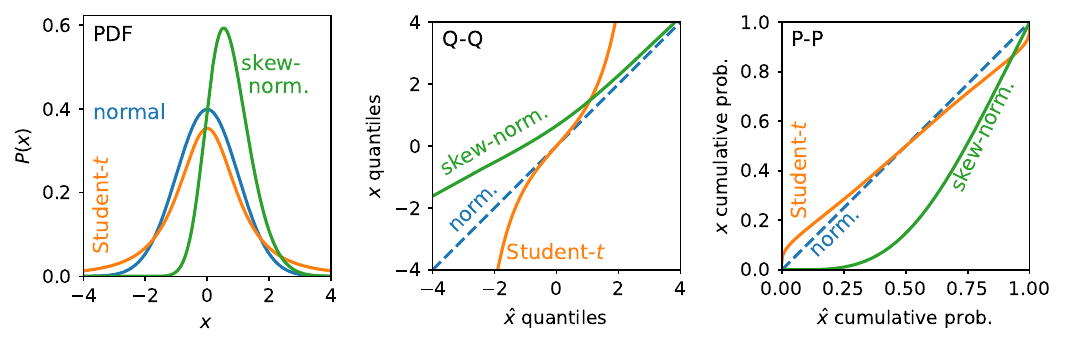}
    \caption{One-dimensional theoretical quantile--quantile (Q--Q) and probability--probability (P--P) plots comparing a normal distribution (blue) to two alternative distributions: a skew-normal (green), and a Student's $t$-distribution (orange).}
    \label{fig:1D}
\end{figure*}

\begin{figure}
    \centering
    \includegraphics[width=0.93\linewidth]{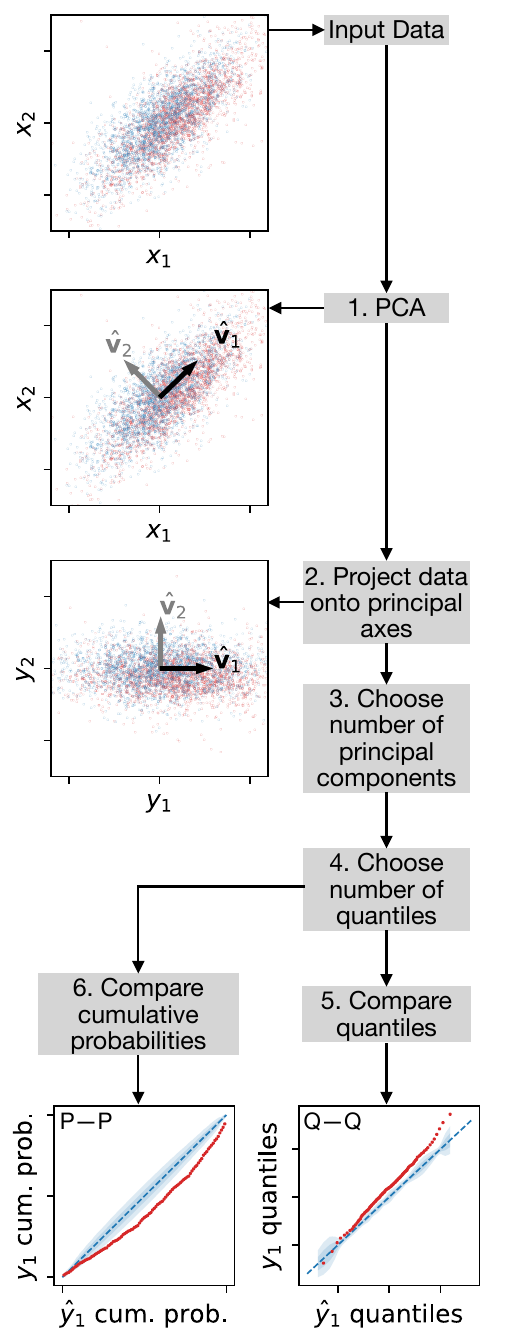}
    \caption{Schematic of the Q--Q and P--P plot-based validation scheme laid out in \S\ref{sec:method}. The steps highlighted in the boxes of the flow chart correspond to the same steps in the mathematical description in Appendix \ref{sec:mathsteps}.}
    \label{fig:flowchart}
\end{figure}

In this paper we present a widely-applicable method for comparing high dimensional samples that uses principal component analysis \citep[PCA;][]{hotelling33a, hotelling33b, hotelling36} for dimensionality reduction (see also \citealp{tortorelli18}), and Q--Q and P--P plots \citep{wilk68} forssment of agreement/disagreement. We also demonstrate a simple bootstrapping approach \citep[following][]{maritz78} for estimating uncertainties, thereby allowing one to verify that any deviations are not due to noise in the quantile estimates. As a numerical summary of the deviation, we use the K--S statistic and the Wasserstein or Mallows distance \citep{kantorovichrubinstein58, vasserstein69, dobrushin70, mallows72, vallender73}; these are intrinsically linked to Q--Q and P--P plots (see \citealp{ramdas17} for an overview), and act as a further compression of the full information shown by the plots.

Our method is well suited to validating and checking models for high-dimensional data directly in the space of the data they are intended to predict. It is particularly suited to identifying the nature of any model mismatches, i.e.,\ where a model is performing sufficiently, and where there are deviations. Conversely, our method is insensitive to model complexity (for a discussion of complexity penalization in model testing, see e.g.\ \S6 of \citealp{ghosh06}). 

Section \ref{sec:method} outlines our proposed methodology, with Section \ref{sec:examples} containing several examples. Section \ref{sec:discussion} includes discussion of possible extensions and generalizations. We conclude in Section~\ref{sec:conclusion}.

\section{Method}
\label{sec:method}
We describe the main steps of our methodology in \S\ref{sec:method_main}, with additional checks detailed in \S\ref{sec:method_additional}.

\subsection{Main Steps}
\label{sec:method_main}

The main ingredients of our proposed validation method are Q--Q and P--P plots, both defined for two sets of samples. A Q--Q plot is constructed by calculating the coordinate values of a fixed number of quantiles for each set of samples, and then plotting the $q$th quantile of one sample against the $q$th quantile of the other. This creates a discrete curve which would show a one-to-one correspondence if the two distributions agreed perfectly. Q--Q plots give greater visual weight to discrepancies in the tails. 

A P--P plot shows the cumulative probability of one sample being below a certain value against the cumulative probability of the other sample being below the same value. This creates a plot bounded between zero and one, which should show a one-to-one correspondence in the case of perfect agreement. A simple one-dimensional example is shown in Figure \ref{fig:1D}. We augment the use of these plots in the case of high-dimensional distributions by using PCA in order to carry out the comparison along a set of principal axes that captures most of the variability in the samples.

The main steps of our method are shown in the flow-chart in Figure \ref{fig:flowchart}. In Appendix \ref{sec:maths} we lay out the exact steps mathematically. Our method can be applied to any situation where one wishes to diagnose the difference between two distributions that are only represented by finite sets of samples (and for which the sample sizes are significantly large compared to the dimensionality of the space they exist in).

We begin by selecting one set of samples --  the data if validating model predictions -- as our reference set, and the other as a test set. We then perform a PCA on the reference sample, and project both samples onto the principal axes of the reference. Next, we choose how many principal components to visualize (e.g.\ the number explaining 90\% of the total variance of the data) and the number of quantiles to compute (see \citealp{hyndman96} for a review of sample quantile estimation). The number of principal components examined should be decided based on the specific analysis setting, and the dimensionality of the uncompressed space\footnote{A scree plot \citep{cattell66} can be used as a heuristic to decide the number of significant components (see Figure \ref{fig:cosmos_scree}).}; caution is needed if the dimensionality is larger than the sample size, where only the principal components with distinctly large variance will be well estimated \citep{jung09}.

We visualize the projections along each axis separately (i.e.\ we do not combine the different axes). In all of the examples we show in Section \ref{sec:examples}, we choose to plot percentiles (i.e.\ the coordinate values that divide the sorted dataset into 100 approximately equal pieces). This choice is made heuristically, as it gives good ``resolution'' of the distributions that are being compared without being excessively noisy in the tails. A Q--Q plot can then be produced by plotting the quantiles of the test sample against the quantiles of the reference sample (see lower right of Figure \ref{fig:flowchart}). A P--P plot is produced by comparing the cumulative probabilities for each sample at a range of coordinate values. This can done by counting the fraction of the test set that is smaller than each quantile of the reference set. The variances of the sample quantiles can be estimated via bootstrapping \citep{maritz78, efron79, bickel81}, allowing the significance of deviations to be visualized.

\subsection{Additional Checks}
\label{sec:method_additional}

There are some pathological situations where dissimilarity could be missed by our approach. For example, consider a reference sample where the 90\% of the total variance is explained by the first $n$ principal components. In a pathological case, the test sample could have marginal distributions that perfectly resemble those of the reference sample when projected along these $n$ axes. But there could be an additional source of variance in the test sample, parallel to the $n+1$'th principal axis of the reference sample; this would be missed. A simple way of flagging such a pathology is to compute the cumulative fraction of variance explained by the selected projections of the reference sample \emph{and} the test sample. If these diverged in a way which was not visible in the Q--Q or P--P plots, it would indicate that additional diagnostics should be used to identify the source of this disagreement.

The method we propose (and many other two-sample tests) would not be directly suitable for situations where the two samples being compared are known to have different noise properties. This would be the case, for example, if noiseless model predictions were being compared to observed data affected by substantial noise. In such situations, it would be desirable to apply an appropriate noise distribution to the model predictions (as is done in forward-modeling pipelines; e.g.\ \citealp{alsing24}) to ensure the comparison is on equal footing.

As a numerical measure of the significance of the discrepancy along each principal axis, the K--S statistic is a natural companion to a P--P plot. This is equal to the largest absolute difference between the empirical distribution function of two samples. This can be directly ``read off'' a P--P plot, corresponding to size of the largest deviation below the diagonal line. To get a sense of the absolute size of the disagreement, a complementary measure is the Wasserstein or Mallows distance between the marginals (for a motivating example, see e.g.\ \citealp{munk98}). In one dimension, the $p$-Wasserstein distance is equal to the distance (using the $L^p$ norm) between two quantile functions (for an introduction see \citealp{levina01, ramdas17, peyre19}). The K--S statistic can be used to test the null hypothesis that the two underlying distributions are equal, giving an associated $p$-value\footnote{The Wasserstein distance (like the Q--Q plot itself; see \citealp{beirlant90}) does not have a distribution-free limiting distribution under this null hypothesis \citep{ramdas17}, so is not straightforwardly associated with a $p$-value.}. Caution should be exercised in the large-data limit, where this test becomes sensitive to increasingly small mismatches that may not be of practical significance \citep[see e.g.][]{hodges54, berger87, lin13, greenland19, gomezdemariscal21}. As such, rejection of the null becomes almost inevitable in applied settings where the model is inherently imperfect \citep[see e.g.][]{berkson38, box76, lindsay09, gelman13}. We thus recommend that the actual size of deviations in the Q--Q and P--P plots, summarised by the K--S statistic and Wasserstein distance are used to contextualize any reported $p$-values.

\section{Examples}
\label{sec:examples}
We now present several demonstrations of the method outlined in Section \ref{sec:method}. We begin with some two-dimensional toy models  (\S\ref{sec:2D} and \S\ref{sec:boomerang}) to give intuition for how different kinds of mismatch are manifested in Q--Q and P--P plots. We follow this with an applied example based on forward modeling galaxy photometric data (\S\ref{sec:cosmos}). Finally, in \S\ref{sec:gmm}, we include a second applied example involving Gaussian mixture models.

\begin{figure*}
    \centering
    \includegraphics[width=\textwidth]{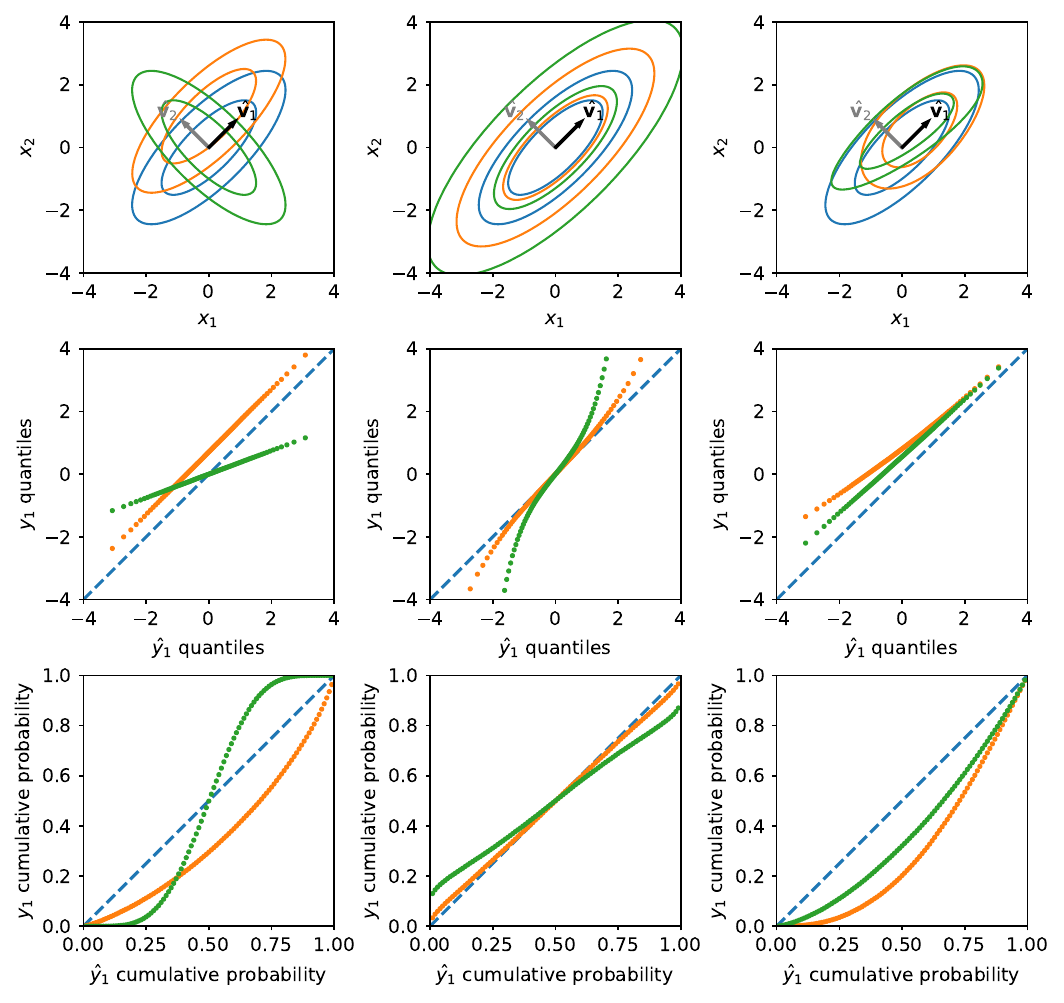}
    \caption{Two-dimensional illustration of quantile--quantile (Q--Q) and probability--probability (P--P) plots along a principal component axis. The upper row shows a variety of bivariate distributions in coordinate space. Our reference distribution (blue) is a multivariate normal in all three columns. In the first column, this is compared to a shifted (orange) and rotated (green) copy of the same distribution. In the second column the alternatives are a multivariate $t$-distribution with five d.o.f.\ (orange), and a multivariate Cauchy (green). In the third column, the alternate distributions are both bivariate skew-normals. The middle row shows Q--Q plots comparing the percentiles of the alternate distributions to the percentiles of the normal reference distribution. The comparison is made along the axis defined by the first principal component of the reference distribution (labelled $\hat{\bm{v}}_1$ in the upper panels). The bottom row shows the equivalent P--P plots.}
    \label{fig:demo}
\end{figure*}

\subsection{Bivariate Normal Distribution}
\label{sec:2D}
First, we will demonstrate the approach on some simple two dimensional examples. We use the bivariate normal distribution as a reference distribution, and compare this to several other multivariate distributions.

Figure \ref{fig:demo} shows an assortment of bivariate distributions being compared to a bivariate normal distribution. The upper row of the figure shows the distributions in coordinate space, with the middle row showing Q--Q plots, and the bottom row showing P--P plots. We show the Q--Q and P--P plots along the first principal component axis of the reference distribution (indicated by the black arrow marked $\hat{\bm{v}}_1$ in the figure). For our reference distribution, we use a bivariate normal distribution with a zero mean vector, unit variance, and a correlation coefficient of 0.75 (shown in blue in all panels of Figure \ref{fig:demo}). The first principal eigenvector along which we will project all of the alternate distributions is $\hat{\bm{v}}_1=(1/\sqrt{2}, 1/\sqrt{2})^\top$.

In the left hand column of Figure \ref{fig:demo} we compare our reference distribution to a shifted and a rotated copy of the same distribution. The shifted distribution (shown in orange) has the same covariance matrix as the reference distribution, but a mean vector of $(0,1)^\top$. This causes a translation away from the diagonal in the Q--Q plot, and a bowing beneath the diagonal in the P--P plot. The rotated distribution (shown in green) has a zero mean vector, unit variance, and a correlation coefficient of $-0.75$. Along the $\hat{\bm{v}}_1$ direction, this is under-dispersed relative to the reference distribution. In the Q--Q plot, under-/over-dispersion manifest as a straight line with a gradient of less/more than unity. In the P--P plot, under-dispersion shows as an S-shaped curve, with over-dispersion being a reflection of this about the diagonal.

In the middle column of Figure \ref{fig:demo} we compare our reference distribution to a pair of heavier tailed distributions (i.e.\ a positive excess kurtosis compared to the normal distribution). In orange we show a multivariate $t$-distribution with 5 degrees of freedom (d.o.f.), whilst in green we show a $t$-distribution with one d.o.f.\ (equivalent to a multivariate Cauchy distribution). For both of these distributions, we use a zero location vector, and scale matrix equal to the covariance of our normal reference distribution. In the Q--Q plots, mismatches of this kind can be seen as a divergence from the diagonal that increases towards the extremes. In the P--P plots, this causes a deviation from the diagonal. Compared to the case of higher variance, the P--P plots for the distributions with excess kurtosis show a much less pronounced turn back towards the diagonal.

In the right hand column of Figure \ref{fig:demo} we compare our reference distribution to a pair of skewed/asymmetric distributions. These are both examples of the bivariate skew normal distribution \citep{azzalini96, azzalini99}. The orange distribution is skewed parallel to $\hat{\bm{v}}_1$, whilst the green has skewness in both directions. In the Q--Q plots, a skewness of this kind can be seen as a translation away from the diagonal, and a slight curvature. In the P--P plot, the skewness creates a bow shape that is less symmetric than the one created by a shift in mean (compare the oranges curves in the lower left and right panels of Figure \ref{fig:demo}).

\subsection{Boomerang Distribution}
\label{sec:boomerang}

\begin{figure*}[t!]
    \centering
    \includegraphics[width=\linewidth]{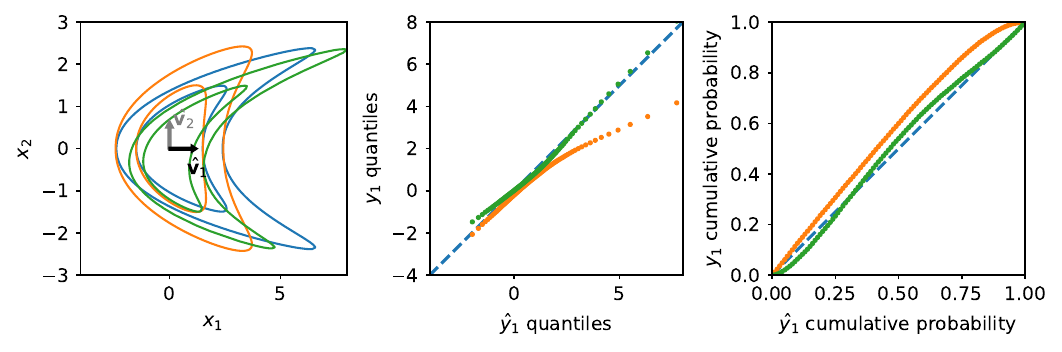}
    \caption{Same as Figure \ref{fig:demo}, but for a ``boomerang'' shaped distribution. We only compare along the first principal axis ($\hat{\bm{v}}_1$) here, as the marginals along the second axis are identical by construction.}
    \label{fig:boomerang}
\end{figure*}

In Figure \ref{fig:boomerang}, we show a more complex two dimensional example. Our reference distribution (blue), and alternative distributions (orange and green) are all  ``boomerang'' transformations of a normal base density. For $\bm{z}=(z_1,z_2)^\top\sim N(\bm{0},\bm{\Sigma})$, we apply a transform given by
\begin{align}
    x_2 &= \theta_2z_2 + \theta_1,\\
    x_1 &= \theta_5z_1 +\theta_3\exp[\ln(|z_2|)\times\ln(1+e^{\theta_4})],
\end{align}
for a parameter vector $\bm{\theta} = (\theta_1,\dots,\theta_5)^\top$. 

For the blue reference distribution and green alternate distribution in Figure \ref{fig:boomerang}, we set $\bm{\theta}=(0,1,1,2,1)^\top$. For the orange alternate distribution, we modify $\theta_3\to0.5$. For blue and orange, we set the covariance matrix of the base distribution to the identity matrix, $\bm{\Sigma}=\bm{I}$. For green, we introduce set the off-diagonal term in $\bm{\Sigma}$ to 0.7 to introduce an asymmetry. The principal axes are parallel to the coordinate axes, and we only show the first principal component $\hat{\bm{v}}_1$ in the Q--Q and P--P plot since the marginals in the $\hat{\bm{v}}_2$ direction are all identical by construction. For the orange distribution, the behaviour is close to the blue in the left tail, but diverges in the right tail due to the shorter ``horns''. The divergence between the blue and green distributions varies in size and direction, with the divergences in the tails better highlighted by the Q--Q plot (middle panel), and the discrepancy in the central part showing more clearly in the P--P plot (right).

\subsection{Forward-modeling Photometric Galaxy Catalogs}
\label{sec:cosmos}

\begin{figure}
    \centering
    \includegraphics[width=\linewidth]{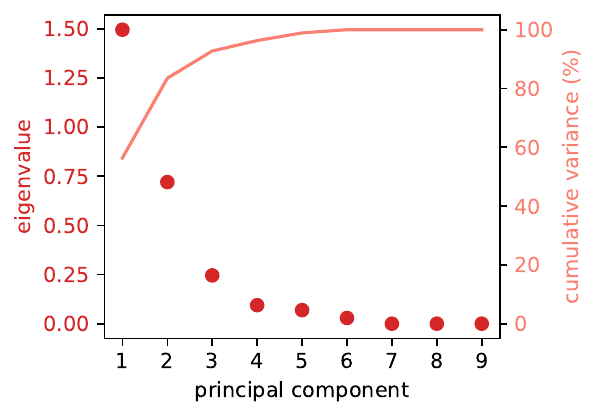}
    \caption{Scree plot showing the eigenvalues of the COSMOS colors' covariance matrix on the left axis, and the cumulative variance explained by the corresponding eigenvectors on the right axis.}
    \label{fig:cosmos_scree}
\end{figure}

\begin{figure*}
    \includegraphics[width=\textwidth]{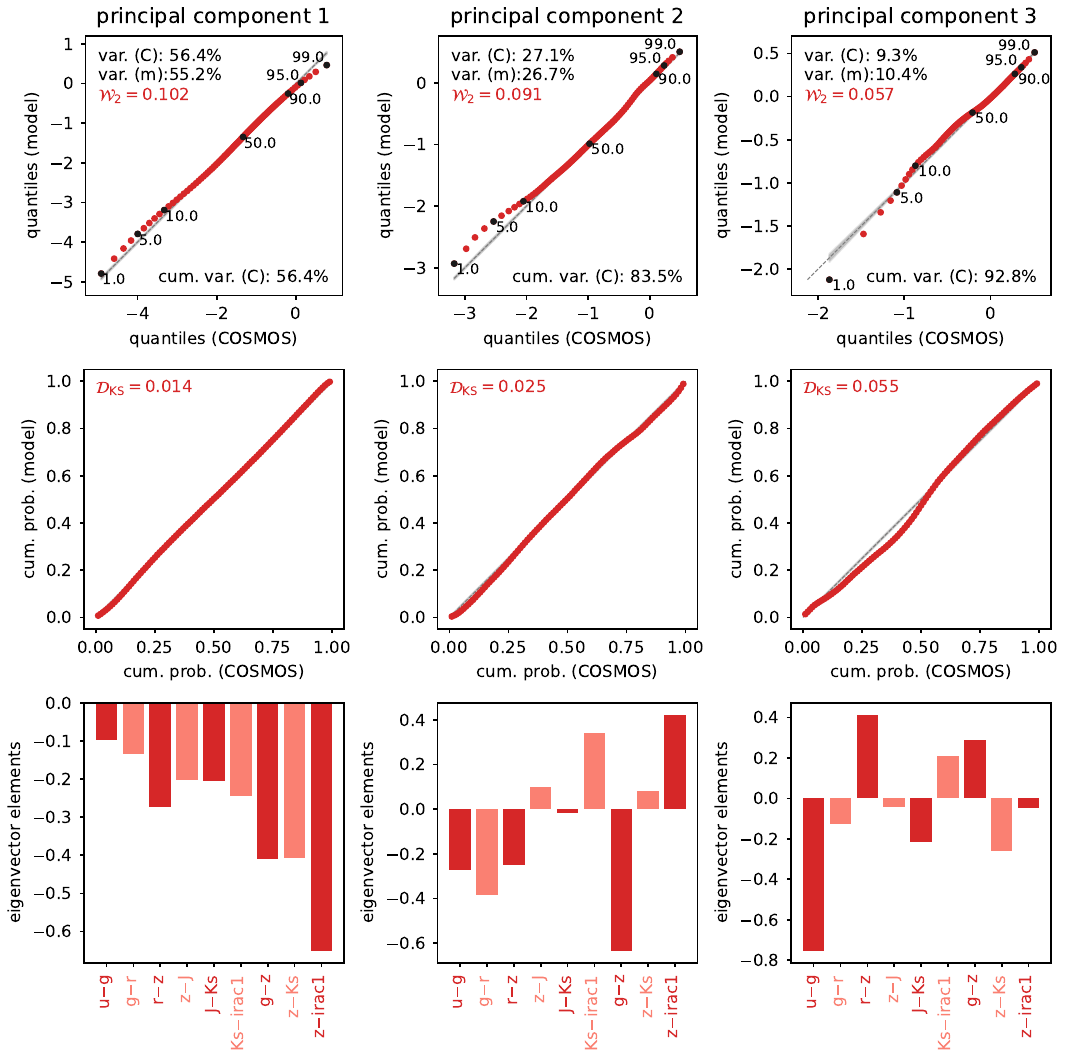}
    \caption{Quantile--quantile (QQ; top row) and probability--probability (PP; second row) plots comparing predictions from the \texttt{pop-cosmos} model (\citetalias{alsing24}) to data from the COSMOS catalog \citep{weaver22}. Comparisons are shown for the first three principal axes of a set of nine broadband colors: $u-g$, $g-r$, $r-z$, $z-J$, $J-K_s$, $K_s-\texttt{irac1}$, $g-z$, $z-K_s$, and $z-\texttt{irac1}$. The contribution of each color to each principal eigenvector is shown in the bar chart in the bottom row of the figure. In the upper left corner of the Q--Q plots, we print the fraction of the total variance in the COSMOS data (C) and model colours (m) explained by the projection along that axis. The first principal component is in the left hand column, which explains around 56\% of the total variance in the COSMOS data. The projection of the model colours along the same axis explains 55\% of the total variance in the model colors. In the lower right corners of the $i$th Q--Q panel, we print the cumulative fractional variance of the COSMOS data explained by principal components 1 through $i$. Cumulatively, the three components shown explain 93\% of the COSMOS data's total variance. On the Q--Q plots, we show the estimated $\mathcal{W}_2$ distance between the model and COSMOS, whilst on the P--P plots we show $\mathcal{D}_\mathrm{KS}$.}
    \label{fig:pop_cosmos_broadband}
\end{figure*}

In a companion paper by \citet[][hereafter \citetalias{alsing24}]{alsing24}, we use the methods developed by \citet{alsing20, alsing23} and \citet{leistedt23} to construct a stellar population synthesis (SPS)-based forward model for the COSMOS photometric galaxy catalog \citep{weaver22}. We will hereafter refer to this model as \texttt{pop-cosmos}. The \texttt{pop-cosmos} model is trained to predict a distribution of magnitudes in 26 of the bands used in the COSMOS survey\footnote{The bands used are: $u$ from the Canada--France--Hawaii Telescope's MegaPrime/MegaCam; $g$, $r$, $i$, $z$, and $y$ from Subaru Hyper Suprime-Cam; $Y$, $J$, $H$, and $K_s$ from UltraVISTA; \texttt{irac1} (Ch1) and \texttt{irac2} (Ch2) from \textit{Spitzer} IRAC; and a set of twelve intermediate and two narrow bands from Subaru Suprime-Cam (IB427, IB464, IA484, IB505, IA527, IB574, IA624, IA679, IB709, IA738, IA767, IB827, NB711, NB816). Full details can be found in \citet{weaver22}.}, based on a learned distribution of galaxy properties. This population distribution is represented by a score-based diffusion model \citep{song19, song20b, song20a, song20j} over 16 physical parameters that closely follow the parameterization used by \texttt{Prospector}-$\alpha$ \citep{leja17, leja18, leja19, leja19_sfh}. These physical parameters are translated into predictions of magnitudes in the 26 COSMOS bands by using an SPS emulator \citep[\texttt{Speculator};][]{alsing20} and a flexible noise model based on a mixture density network \citep{bishop06}. Training of the \texttt{pop-cosmos} model is carried out by minimizing the Wasserstein distance (approximated using the Sinkhorn divergence; \citealp{cuturi13, feydy19}). In this section we use our proposed validation framework to compare predictions from the \texttt{pop-cosmos} model to the COSMOS data themselves. Since the realization of noise is included within the generative model, we will be comparing predicted \emph{noisy} magnitudes directly to the COSMOS data. We therefore do not use any special strategy to account for the noise in the data, as predicting realistic noise is a key step of the generative process. Possible extensions of our methodology to handle noisy data and noiseless model predictions are discussed in \S\ref{sec:discussion}.

\begin{figure}[ht!]
    \centering
    \includegraphics[width=\linewidth]{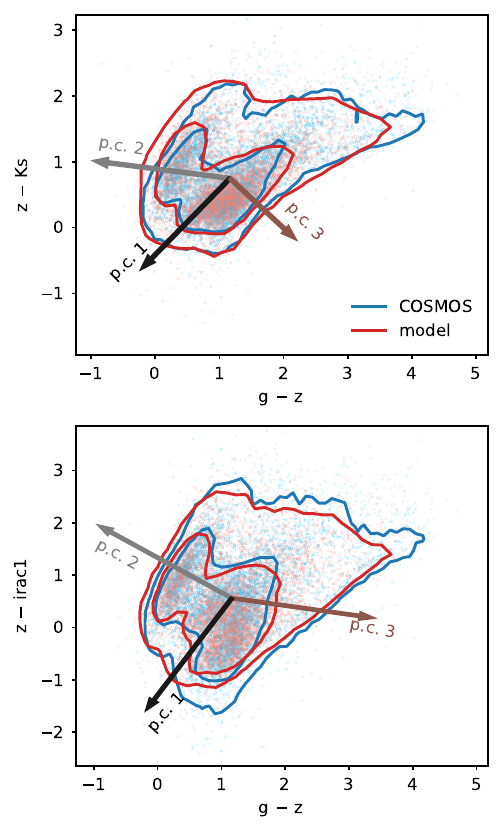}%
    \caption{Color--color diagrams ($z-K_s$ and $z-\texttt{irac1}$ vs.\ $g-z$; c.f.\ figure 12 of \citealp{weaver22}). The first, second, and third principal eigenvectors of the COSMOS data are shown as black, grey, and brown arrows respectively. The light blue and light red points show the colors of 7,000 randomly selected galaxies from COSMOS ($\approx5$\% of the sample analysed in \citetalias{alsing24}) and the \texttt{pop-cosmos} model. Contours are computed from the full sample, with the inner contours containing 68\% of galaxies, and the outer contours containing 95\%.}
    \label{fig:colors12}
\end{figure}

As detailed fully in \citetalias{alsing24}, we apply an $r$-band magnitude cut of $r<25$ to both the data (leaving $\approx140,000$ galaxies) and the model predictions and then compute an array of broadband colors for both sets of magnitudes. In particular, we consider the following nine colors: $u-g$, $g-r$, $r-z$, $z-J$, $J-K_s$, $K_s-\texttt{irac1}$, $g-z$, $z-K_s$, and $z-\texttt{irac1}$. The later three are highlighted in figure 12 of \citet{weaver22}, whilst the remainder are highlighted in their figure 9 as being of particular importance to constraining galaxy properties via SED modeling.

\begin{figure}[ht!]
    \centering
    \includegraphics[width=\linewidth]{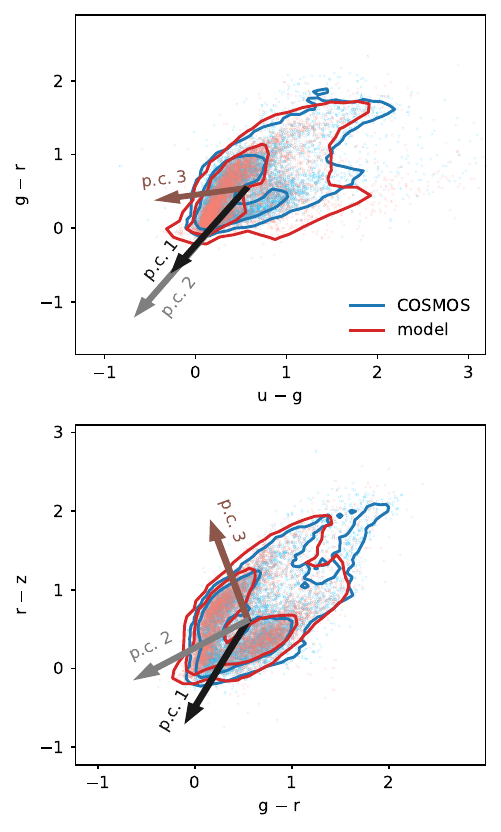}
    \caption{Same as Figure \ref{fig:colors12}, but showing $g-r$ vs.\ $u-g$, and $r-z$ vs.\ $g-r$.}
    \label{fig:colors23}
\end{figure}

We show a scree plot \citep{cattell66} of the sorted eigenvalues of the covariance matrix of the COSMOS sources' colors in Figure \ref{fig:cosmos_scree}. This shows that the first three principal components are the most significant, together contributing over 90\% of the variance. Figure \ref{fig:pop_cosmos_broadband} shows Q--Q and P--P plots for the first three principal components of our chosen list of colors. These three principal components cumulatively explain around 93\% of the total variance in the COSMOS colors, with the first explaining around 56\%. e check that the (cumulative) fraction of total variance explained by this projection of the model colors is comparable. In this space, we compare the predicted color distributions from \texttt{pop-cosmos} (along the vertical axes in the top two rows) to the COSMOS data (along the horizontal axes of the top two rows). Due to the high number of objects, the variance of the sample quantiles is small, so the uncertainty band is very narrow. The K--S statistic $\mathcal{D}_\mathrm{KS}$ between the two samples is relatively small. The $\mathcal{W}_2$ distance is also small along all axes, agreeing with the visual impression that the model and data are in good agreement. The bottom row of Figure \ref{fig:pop_cosmos_broadband} shows the contributions of the nine colors to each of the principal eigenvectors; this visualization serves to add interpretability to the findings. 

The first principal component contains significant contributions from the full list of colors, weighted towards towards the three more widely-spaced colors ($g-z$, $z-K_s$, and $z-\texttt{irac1}$). The agreement seen between the \texttt{pop-cosmos} model and the data is visually very good along this principal axis, with slight deviations in the tails of the distribution.

The second principal component again contains substantial contributions from most of the colors apart from $z-J$, $J-K_s$, and $z-K_s$. Along this principal axis, we see deviation between the model predictions and the data below the 10th percentile -- visible most clearly in the Q--Q plot in the upper centre of Figure \ref{fig:pop_cosmos_broadband}. The agreement for the upper 90\% of galaxies is good. The divergence seen on the edge suggests that the data are heavier tailed in the lower tail of this component. Figure \ref{fig:colors12} shows $z-K_s$ and $z-\texttt{irac1}$ vs.\ $g-z$ for COSMOS and \texttt{pop-cosmos}, with the second principal eigenvector of the COSMOS data shown as a grey arrow (p.c.\ 2). In both panels, it is noticeable that the COSMOS data has a slightly heavier lower tail in this direction (towards the right of the figure). Figure \ref{fig:colors23} shows that the $u-g$, $g-r$, and $r-z$ directions also contribute to this in the same way. Some of this disagreement could be due to difficulties with the COSMOS photometry in the $u$- and $g$-bands, which \citet{weaver22} find exhibit systematic differences between two different photometry catalogs (\texttt{Classic}, comparable to \citealp{laigle16}; and \texttt{Farmer}, based on \citealp{weaver23}). These systematics, and the relevance to the \texttt{pop-cosmos} model are discussed further by \citetalias{alsing24}, who note that systematic offsets between the \texttt{pop-cosmos} model predictions and the COSMOS data are smaller than the offsets between \texttt{Farmer} and \texttt{Classic}.

The third principal axis is dominated by the $u-g$ color, and shows the strongest disagreement between model and data. Both the Q--Q and P--P plots show that the disagreement is more substantial below the median. Inspecting the upper panel of Figure \ref{fig:colors23}, we can see that the model does not fully capture the structure of the data in the space of $g-r$ vs.\ $u-g$. In particular, the model predictions are much more concentrated into one side of the U-shape delineated by the inner blue contour (containing 68 per cent of the COSMOS galaxies). Some of this discrepancy may be explained by potential systematics in the data (discussed above), and the fact that the $u$- and $g$-bands tend to have a lower signal-to-noise and more very faint galaxies (c.f.\ figure 4 in \citetalias{alsing24}).

\subsection{Five-Component Gaussian Mixture}
\label{sec:gmm}
\begin{figure}
    \centering
    \includegraphics[width=\linewidth]{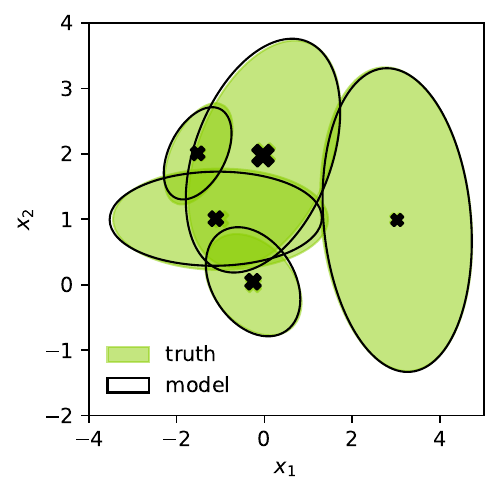}
    \caption{Fitted GMM (black ellipses) and ground truth (green) for the example in \S\ref{sec:gmm}. The crosses mark the means of the components, with size of the crosses being proportional to the weights of the different components.}
    \label{fig:gmm}
\end{figure}

As a final example, we consider a five-component Gaussian mixture model (GMM), motivated by the use of GMMs as an estimator of mutual information\footnote{For an introduction see \citet[\S2.4, 8.5]{cover05}. For example applications in astronomy, see e.g.\ \citet{pandey17, jeffrey21, upham21, chartab23, luciesmith22, luciesmith24mnras, guo24, piras24}.} between two quantities \citep[e.g.][]{piras23}. In our example, the ground truth is a joint distribution over two variables, $\bm{x}=(x_1,x_2)^\top$, given by a five-component GMM as
\begin{equation}
    P(x_1,x_2) = \sum_{c=1}^5 w_c\,N(\bm{x}|\bm{\mu}_c,\bm{\Sigma}_c),
    \label{eq:gmm}
\end{equation}
where $w_c$, $\bm{\mu}_c$, and $\bm{\Sigma}_c$ are respectively the weight, mean, and covariance of the $c$'th component, with the weights subject to the constraint that $\sum_{c=1}^5w_c = 1$. We fit a GMM to a set of samples drawn from $P(\bm{x})$, using the expectation maximization algorithm \citep{dempster77, melchior18}, as in \citet{piras23}. The resultant fit, which is compared to the ground truth in Figure \ref{fig:gmm}, matches sufficiently well for any practical scientific use.

The mutual information between $x_1$ and $x_2$ is defined by
\begin{equation}
    I(x_1, x_2) = \iint_{-\infty}^\infty P(x_1,x_2)\ln\left[\frac{P(x_1,x_2)}{P(x_1) P(x_2)}\right]\,dx_1\,dx_2,
\end{equation}
where $P(x_1,x_2)$ is the joint distribution over $x_1$ and $x_2$ (Eq.\ \ref{eq:gmm}), and $P(x_1)$ and $P(x_2)$ are the marginal densities. For the ground truth, we compute $I(x_1,x_2)=0.180$~nats (where nats refers to natural units) via quadrature integration\footnote{We use the \texttt{QUADPACK} routines \citep{piessens83} via \texttt{SciPy}.}. From the fitted model, we compute $I(x_1,x_2)=0.182\pm0.006$~nats, in excellent agreement and confirming the sufficiency of the fit\footnote{We compute the uncertainty on $I(x_1,x_2)$ via bootstrap, following \citet{piras23}, by resampling with replacement from the original set of samples and re-fitting the GMM 50 times. Note that this is distinct from the bootstrapping procedure used for the Q--Q and P--P plots.}.

\begin{figure}
    \centering
    \includegraphics[width=\linewidth]{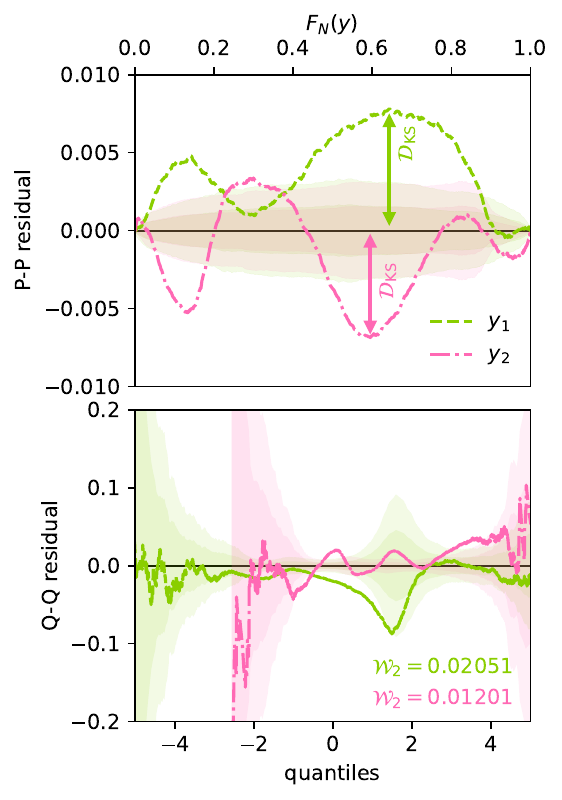}
    \caption{Residuals about the diagonal for P--P (top panel) and Q--Q (bottom panel) plots computed for the GMM fit in \S\ref{sec:gmm}. The P--P residuals show the difference in the empirical distribution functions (for $-\infty<y<\infty$) of $N=10^6$ draws from the fitted model and ground truth. The Q--Q residuals are unbinned, and show the difference between the $n$th largest draws (for $n=1,\dots,10^6$) from the model and truth, plotted at the location of the $n$th largest draw from the true distribution. Shaded regions in both panels show the 68\% and 95\% regions estimated from bootstrapping (\S\ref{sec:bootstrap}).}
    \label{fig:gmm_residuals}
\end{figure}

We take $N=10^6$ draws from the fitted model and the true GMM, project these along the two principal component axes (which are almost parallel to the axes in Figure \ref{fig:gmm} in this case), and compare their quantiles and cumulative probabilities. Figure \ref{fig:gmm_residuals} shows the deviations of the resulting Q--Q and P--P plots about the diagonal line, with green and pink corresponding to the two axes. The P--P and Q--Q plots both show small deviations about the diagonal, indicative of the small deviations from the truth in the shapes of the fitted GMM components. The deviations in the P--P plot are noticeably larger than the uncertainty envelope estimated via bootstrapping, as the large sample size resolves the empirical distribution functions precisely enough for such small differences ($<0.01$ for the full domain of $[0,1]$) to be confidently identified. For this example, we show the residuals from an `unbinned'' Q--Q plot; rather than selecting a fixed number of quantiles, this is obtained by simply sorting the two samples and plotting these directly against one another (giving maximum resolution in the center of the distributions, but larger uncertainty towards the tails). Even in the tails of the distributions, the quantiles differ by $\lesssim0.1$, which is very small compared to the full range spanned by the distributions.

We highlight in Figure \ref{fig:gmm_residuals} the correspondence of $\mathcal{D}_\mathrm{KS}$ to the largest deviation in the P--P plot.  Following e.g.\ \citet{levina01}, the 2-Wasserstein distance $\mathcal{W}_2$ is given directly by the root mean square of the deviations in the Q--Q residual curve (see details in \S\ref{sec:metrics}). For the first prinicpal component, we calculate a Wasserstein distance of $\mathcal{W}_2\approx0.02$, showing that the integrated difference in the quantile functions is very small. We calculate $\mathcal{D}_\mathrm{KS}\approx0.008$, corresponding to a K--S test $p$-value of $\sim10^{-27}$ for $N,M=10^6$. This implies an extremely high-confidence rejection of the null hypothesis that the fitted model and truth have the same distribution function. This is consistent with the impression from the bootstrap; with such a large sample size, it is easy to unambiguously identify the extremely small deviations between the fitted GMM and the truth. However, as we have seen with the mutual information calculation, these deviations are not large enough to limit the usefulness of the fitted model.

\begin{figure}
    \centering
    \includegraphics[width=\linewidth]{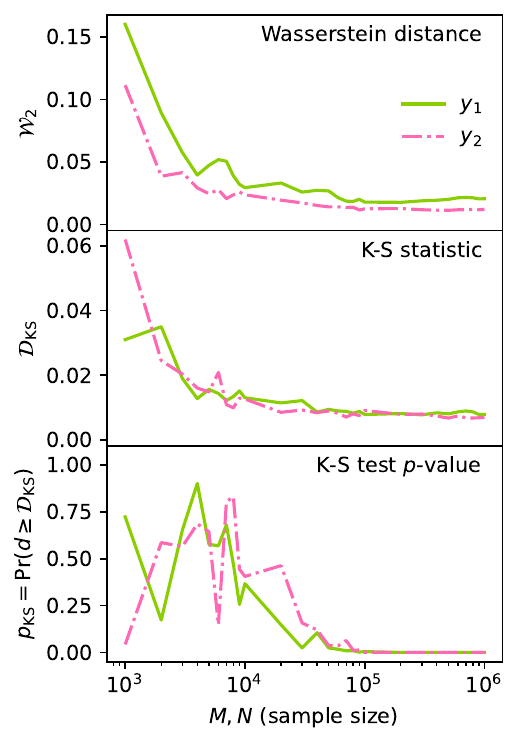}
    \caption{Sample-size dependence of $\mathcal{W}_2$ (top panel), $\mathcal{D}_\mathrm{KS}$ (middle panel), and the K--S test $p$-value $\mathrm{Pr}(d\geq\mathcal{D}_\mathrm{KS})$ (bottom panel) for the example in \S\ref{sec:gmm}. The green curves show the metric computed along the first principal component direction of the $\bm{x}$ distribution, with the pink showing the second principal component direction.}
    \label{fig:gmm_metrics}
\end{figure}

To further explore the behavior of the summary metrics, we take  varying numbers (from $N=10^3$ to $N=10^6$) of draws from the ground truth and fitted model, and use these to compute $\mathcal{D}_\mathrm{KS}$ and $\mathcal{W}_2$ between the fitted model and the truth. Figure \ref{fig:gmm_metrics} shows how $\mathcal{D}_\mathrm{KS}$ (middle panel) and $\mathcal{W}_2$ (top panel) vary with sample size. For each sample size, we take the same number of draws from the fitted model and the truth. We show the metrics for the two principal component directions, with the behaviour being very similar for both.

In Figure \ref{fig:gmm_metrics}, we can see that for the first principal component, the K--S statistic converges towards $\mathcal{D}_\mathrm{KS}\approx0.008$ as the sample size increases, with the Wasserstein distance tending towards $\mathcal{W}_2\approx0.02$. These values are reflective of the actual difference between the cumulative distribution and quantile functions of the ground truth and the (slightly imperfect) fitted model; as the sample size grows, our ability to estimate the true size of the difference becomes increasingly precise.

By contrast, the lower panel of Figure \ref{fig:gmm_metrics} shows the K--S test $p$-value; even though the estimated size of $\mathcal{D}_\mathrm{KS}$ remains constant for $N\gtrsim 10^5$, the associated $p$-value becomes arbitrarily small, as we have already seen above. This is indeed the expected behaviour (see e.g.\ extensive discussion in \citealp{greenland19}), reflecting our ability to confidently identify arbitrarily small differences between the distributions of the two samples given sufficiently high $N$. This highlights the inherent difficulty with using a $p$-value as a decision threshold in the large-$N$ regime for any applied modeling problem: with enough data, it will always be possible to identify a statistically significant discrepancy. One must therefore inspect the size and nature of any differences between a model's predictions and the data, which is possible in the more nuanced visual testing approach we advocate.
\vfill
\section{Discussion}
\label{sec:discussion}
We now discuss related work, and possible future extensions to our methodology. We present in \S\ref{sec:wasserstein} additional discussion of the Wasserstein distance, and related metrics and statistical tests. In \S\ref{sec:nonlinear}, we discuss the possibility of using non-linear projections as a generalization of PCA. In \S\ref{sec:hdlss}, we discuss the limit of high dimensionality relative to sample size. Finally, in \S\ref{sec:misc}, we briefly highlight probabilistic PCA, and other possible algorithmic improvements.

\subsection{Wasserstein and Related Metrics}
\label{sec:wasserstein}
In this work, we have opted to use the K--S statistic and 2-Wasserstein distance to add a numerical summary of the discrepancy along each projected axis.  Whilst these are quantitative, they encode less information about any potential model mismatch than the plots from which they are derived. The K--S statistic is equal to the size of the largest deviation shown in a P--P plot\footnote{In fact, the deviations in a P--P plot converge towards a Brownian bridge \citep[e.g.][]{doksum74, aly87, beirlant90}, the supremum of which defines the Kolmogorov distribution (the limiting distribution of the K--S statistic; for further background see \citealp{hajek99}, ch.\ 6).}. The Wasserstein distance is directly related to the distance between quantile functions \citep{vallender73, bickel81}; it is an integral metric like the K--S statistic \citep[see e.g.][]{muller97}; and it can be formally connected to many other two-sample tests (for a review see \citealp{ramdas17}).

Various other metrics and statistical hypothesis tests based on the Wasserstein distance have been developed to handle truncated/trimmed distributions \citep[e.g.][]{munk98, freitag07, alvarezesteban08, alvarezesteban12}, or one-sample testing against a known distribution \citep[e.g.][]{cuestaalbertos99, delbarrio00}. A quantile-based two-sample test is also proposed independently by \citet{kosorok99} for the specific setting of comparing survival functions estimated from incomplete data in clinical trials \citep[e.g.][]{kaplan58}. The ``energy distance'' \citep[e.g.][]{baringhaus04, szekely04, szekely05, szekely13}, which is a commonly used multivariate discrepancy measure, is shown by \citet{ramdas17} as being related to the Wasserstein distance with entropic smoothing \citep{cuturi13}. Kernel-based test statistics, such as the maximum mean discrepancy \citep[e.g.][]{gretton06, gretton08, gretton12}, have also been shown to be related to the energy and Wasserstein distances \citep[see][]{sejdinovic13, ramdas17}. In one dimension (i.e.\ when comparing marginals as we have in this work), the Wasserstein distance is inexpensive. In a multivariate setting, it is  considerably more costly when using an exact approach (e.g.\ \citealp{kuhn55, munkres57}; for an extensive review see \citealp{peyre19}), but can be made more tractable using the \citet{sinkhorn67} algorithm of \citet{cuturi13}. However, comparing the marginals of two distributions using the Wasserstein distance is typically sufficient for assessing similarity \citep{puzicha99, rubner01, levina01}.

\subsection{Non-linear Dimensionality Reduction}
\label{sec:nonlinear}
In situations where the distributions being compared show very strong non-linearities or curving degeneracies, it may also be appropriate to consider a generalization of PCA that uses a non-linear projection of the data (e.g.\ principal curves, kernel PCA, or locally linear embedding; \citealp{hastie89, tibshirani92, scholkopf98, mika98, roweis00}; see also discussion in \citealp{bishop06}).

Non-linear extensions to PCA have occasionally been used to identify structure in astronomical datasets \citep[e.g.][]{einbeck07, vanderplas09, taghizadeh12, bu13, ward16}, but they have not been widely tested in the field. More sophisticated non-linear projections such as variational autoencoders (VAEs) have also been employed more recently for dimensionality reduction for astronomical data \citep[e.g.][]{portillo20}. The combination of quantile comparison with a VAE (or similar) would be an interesting avenue for future research. A trained VAE could be use to compress two or more samples down to a low-dimensional latent space, after which a quantile-based comparison could be performed to test for consistency of the two samples. Given the recent breakthroughs in the use of VAEs to find a physically meaningful latent representation of complex physical systems \citep[see e.g.][]{higgins17, iten20, luciesmith22, luciesmith24prl, luciesmith24mnras, akhmetzhanova24, guo24}, this would be worth exploring. However, the greater computational cost involved in training a VAE, and the challenge of constructing a meaningful and decorrelated latent space, would make this less immediately or generically applicable than a simple compression like PCA.

Another approach that could be explored in future astrophysical applications is the use of space-filling curves \citep[e.g.][]{skilling04a} to map a $D$-dimensional space to a one dimensional space where univariate statistical tests can be applied \citep{heuchenne22}. Such a transformation has previously been used \citep{skilling04b} to enable an efficient variant of slice sampling \citep{neal03}, and as a novel approach to computing the Wasserstein distance \citep{bernton19}. A generalization of our current methodology to make use of the connections highlighted by \citet{bernton19} and \citet{heuchenne22} would be a fruitful direction for further research.

\subsection{High Dimension, Low Sample Size Data}
\label{sec:hdlss}
The motivating example we have considered in this work involves sample sizes which are large relative to the dimensionality of the space (i.e.\ $D\ll N$), and we expect our proposed methodology to be appropriate primarily in this setting. In the opposite regime (i.e.\ $N\ll D$; for a conceptual review see \citealp{aoshima17} or \citealp{aoshima18}), the pairwise distance between randomly sampled vectors in the $D$-dimensional space tends towards a constant as $D\to\infty$ \citep[see e.g.][]{hall05, ahn07}, and the geometry of a set of $N$ points becomes highly constrained. In such a setting, the eigenvectors of the sample covariance matrix (i.e.\ the estimated principal component directions) are only reliable estimators of the true principal components if they correspond to distinctly large eigenvalues \citep[see e.g.][]{jung09, johnstone09, yata09, yata13}. \citet{takeuchi24} recently presented an analysis of astronomical data in the $N\ll D$ limit; specifically the Atacama Large Millimeter Array (ALMA) spectroscopic map of NGC~253. In this setting, \citet{takeuchi24} test the use of several ``sparse'' forms of PCA \citep{yata10, yata12, yata22} that are expected to yield more consistent principal eigenvectors in this setting (see also \citealp{johnstone09, yata13, shen13}). Such extensions of PCA could potentially be employed to scale our methodology to the $N\ll D$ setting, although we expect that the usefulness of a quantile-based comparison would be limited for small $N$.

The performance of classical PCA has been shown to be excellent in very high-dimensional settings when $N$ is sufficiently large. A particularly noteworthy success in the astronomy literature has been the compression of large samples of spectra \citep[e.g.][]{bailer98, yip04, wild05, hu12, patil22} or spectral models \citep[e.g.][]{han14, czekala15, alsing20} using PCA for a variety of different object classes. PCA and related techniques have shown success in astrophysical applications even in the $N\lesssim D$ limit \citep[e.g.][]{mittaz90, connolly95, sodre97, formiggini04, wang11}, as well as in other domains (see examples in \citealp{aoshima18}).

\subsection{Algorithmic Improvements}
\label{sec:misc}
In some situations a probabilistic formulation of PCA, where noise information can be included in the statistical model via the likelihood function \citep[e.g.][]{roweis97, tipping99, wentzell99, wentzell12, bailey12, chen22}, may lead to a more optimal projection than traditional PCA. Another possible future development could be the adoption of QR-decomposition PCA \citep{sharma13, desouza22} for improved scaling over covariance or singular-value decomposition (SVD) approaches. Further investigation of the performance of our uncertainty quantification scheme (\S\ref{sec:bootstrap}), including the consideration of alternative estimators \citep[e.g.][]{bloch68} for the variance of sample quantiles (see \citealp{hall88}), may also be worthwhile.

\section{Conclusions}
\label{sec:conclusion}
We have presented a simple model checking and validation scheme that can be used for comparison of two high-dimensional samples (e.g.\ model predictions and data). These situations are ubiquitous in astrophysics and cosmology -- especially in the context of generative machine learning (ML). Our proposed approach is based on performing PCA on one of the samples, typically the data, and then comparing the two samples along the principal axes of the data. The comparison is made using quantile--quantile (Q--Q) and probability--probability (P--P) plots.  We have demonstrated our methodology on a set of simple two-dimensional examples (Figure \ref{fig:demo} and \ref{fig:boomerang}), to provide intuition for how different kinds of model mismatch are manifested in the Q--Q and P--P plots. We have then successfully applied our methodology to the validation of an ML-based generative model for galaxy photometry (\texttt{pop-cosmos}; \citetalias{alsing24}).

We have demonstrated  that the use of Q--Q and P--P plots in combination with PCA can be an effective way to augment other more commonly used data space comparison techniques (e.g.\ comparing cumulative distribution functions, inspecting one- and two-dimensional marginal distributions). Although we have been motivated by validation of ML-based population models, we expect that the technique we have presented can be applied more broadly in other settings where two or more sets of samples from (potentially unknown) distribution are being compared.

\vfill
\noindent\textbf{Author contributions.} 
We outline the different contributions below using keywords based on the CRediT (Contribution Roles Taxonomy) system. 
\textbf{ST:} conceptualization, methodology, formal analysis, visualization, writing -- original draft, writing -- editing \& review. {\bf HVP:} conceptualization, methodology, validation, visualization, writing -- editing \& review, supervision, funding acquisition. {\bf DJM:}  conceptualization, methodology, validation, visualization, writing -- editing \& review. {\bf JA:} data curation, validation, writing -- editing \& review. {\bf BL:} methodology, writing -- editing \& review. {\bf SD:} writing -- editing \& review. 

\noindent\textbf{Data availability.}
The \texttt{pop-cosmos} model draws in \S\ref{sec:cosmos} will be shared on reasonable request to the corresponding author. The COSMOS catalog \citep{weaver22} used in \S\ref{sec:cosmos} is publicly available at the \dataset[ESO Archive]{https://doi.org/10.18727/archive/52} and the \dataset[COSMOS2020 webpage]{https://cosmos2020.calet.org}.

\noindent\textbf{Acknowledgments.} 
We thank James Alvey, Anik Halder, George Efstathiou, Joel Leja, and Arthur Loureiro for useful comments and discussions, and Davide Piras for providing a toy problem  studied in this work. We also thank the reviewer and editors at ApJS for their constructive and  helpful comments that lead to an improved manuscript. ST, HVP, JA, and SD have been supported by funding from the European Research Council (ERC) under the European Union's Horizon 2020 research and innovation programmes (grant agreement no. 101018897 CosmicExplorer). This work has been enabled by support from the research project grant ‘Understanding the Dynamic Universe’ funded by the Knut and Alice Wallenberg Foundation under Dnr KAW 2018.0067. HVP was additionally supported by the G\"{o}ran Gustafsson Foundation for Research in Natural Sciences and Medicine. HVP and DJM acknowledge the hospitality of the Aspen Center for Physics, which is supported by National Science Foundation grant PHY-1607611. The participation of HVP and DJM at the Aspen Center for Physics was supported by the Simons Foundation. This research also utilized the Sunrise HPC facility supported by the Technical Division at the Department of Physics, Stockholm University.  BL is supported by the Royal Society through a University Research Fellowship.

%


\software{\texttt{NumPy} \citep{harris20}; \texttt{SciPy} \citep{virtanen20}; \texttt{Matplotlib} \citep{hunter07}; \texttt{PyTorch} \citep{paszke19}; \texttt{scikit-learn} \citep{pedregosa11}; \texttt{GMM-MI} \citep{piras23}}
%

%
\appendix
\section{Mathematical Description of the Testing Algorithm}
\label{sec:maths}
\subsection{Mathematical Outline}
\label{sec:mathsteps}
A full mathematical description of our proposed validation algorithm from Section \ref{sec:method} is provided here. Our inputs are two sets of samples of $D$-dimensional quantities (i.e.\ two $D$-dimensional point clouds), drawn from potentially unknown distributions. One of these sets of observations (of size $N$) is our reference distribution, which we will be comparing the other set (of size $M$) to. The $n$th observation in the reference set is a $D$-dimensional vector $\hat{\bm{x}}_n$, where hats are used because this is typically a sample of observed/measured values. The $m$th sample in the alternate set is a vector $\bm{x}_m$. The reference observations are combined into an $N\times D$ matrix $\hat{\bm{X}}$, whose $n$th row is $\hat{\bm{x}}_n^\top$. Similarly, we construct an $M\times D$ matrix $\bm{X}$, whose $m$th row is $\bm{x}_m^\top$. The algorithm is best suited to the case that $D\ll N,M$; for $D \gtrsim M,N$, an SVD-based PCA is more stable (step \ref{step:SVD}; see \citealp{mittaz90} for discussion in an astronomy context).

With the inputs assembled, we proceed as follows:
\begin{enumerate}
    \item Perform a PCA on the data:
    \begin{enumerate}
        \item Compute the sample mean of the reference sample, $\hat{\bm{\mu}} = N^{-1}\sum_{n=1}^N\hat{\bm{x}}_n$;
        \item Compute the $D\times D$ covariance matrix of the reference sample, $\hat{\bm{C}}=(N-1)^{-1}\sum_{n=1}^N(\hat{\bm{x}}_n-\hat{\bm{\mu}})(\hat{\bm{x}}_n-\hat{\bm{\mu}})^\top$;
        \item \label{step:eig} Take an eigen-decomposition, $\hat{\bm{C}}=\hat{\bm{V}}\hat{\bm{\Lambda}}\hat{\bm{V}}^{-1}$, where $\hat{\bm{\Lambda}} = \diag(\hat{\bm{\lambda}})$, $\hat{\bm{\lambda}} = (\hat{\lambda}_1,\dots,\hat{\lambda}_D)^\top$ is a vector containing the $D$ eigenvalues of $\hat{\bm{C}}$ ordered from largest to smallest, and $\hat{\bm{V}}$ is a $D\times D$ matrix whose $d$th column $\hat{\bm{v}}_d$ contains the eigenvector corresponding to $\hat{\lambda}_d$.
        \item \label{step:SVD} (Alternatively, for $D\gtrsim N$) Take a singular-value decomposition of the mean-centred data matrix, $\hat{\bm{X}} - \hat{\bm{\mu}}^\top = \hat{\bm{U}}\hat{\bm{\Sigma}}\hat{\bm{V}}^\top$, where $\hat{\bm{U}}$ and $\hat{\bm{V}}$ are respectively $N\times N$ and $D\times D$ matrices whose columns are orthogonal unit vectors. The diagonal $N\times D$ matrix $\hat{\bm{\Sigma}}$ obeys $\hat{\bm{\Sigma}}^\top\hat{\bm{\Sigma}} = \hat{\bm{\Lambda}}$, where $\hat{\Lambda}_{dd} = \hat{\lambda}_d$ is the $d$th eigenvalue from step \ref{step:eig} (i.e.\ $\hat{\Sigma}_{dd}=\hat{\lambda}_d^{1/2}$). As in step \ref{step:eig}, the columns of $\hat{\bm{V}}$ are the eigenvectors of interest, with the $d$th column corresponding to the eigenvalue $\hat{\lambda}_d=\hat{\Sigma}_{dd}^2$.
    \end{enumerate}
    \item Project both samples onto the principal axes defined by $\hat{\bm{V}}$, by computing $\hat{\bm{Y}} = \hat{\bm{X}}\hat{\bm{V}}$ and $\bm{Y} = \bm{X}\hat{\bm{V}}$. The $d$th column of $\hat{\bm{Y}}$, $\hat{\bm{y}}_d$, contains the projection of all observations onto the axis defined by $\hat{\bm{v}}_d$.
    \item Choose the number of principal components $D_\mathrm{max}$:
    \begin{enumerate}
        \item Create a scree plot of $\hat{\lambda}_d$ vs.\ $d$ for $d=1,\dots,D$.
        \item For a chosen fraction $f$ of the total variance, find the smallest $D_\text{max}$ satisfying $f\sum_{d=1}^{D}\hat{\lambda}_d\geq\sum_{d=1}^{D\text{max}}\hat{\lambda}_d$.
    \end{enumerate} 
    \item\label{step:chooseQ} Choose $Q$, the number of quantiles to be computed. We define the $Q$ quantiles of a dataset as the coordinate values that partition the sorted dataset into $Q+1$ almost evenly sized pieces (\S\ref{sec:quantiles}).
    \item\label{step:QQ} For $d=1,\dots,D_\text{max}$, create a Q--Q plot:
    \begin{enumerate}
        \item Extract the $d$th columns, $\hat{\bm{y}}_d$ and $\bm{y}_d$, from $\hat{\bm{Y}}$ and $\bm{Y}$, respectively;
        \item\label{step:quantiles} For $q=1,\dots,Q$, calculate the $q$th quantile of $\hat{\bm{y}}_d$ and $\bm{y}_d$, respectively $\hat{y}_d^q$ and $y_d^q$ (\S\ref{sec:quantiles});
        \item Plot $\hat{y}_d^q$ vs.\ $y_d^q$ for all $q$;
        \item\label{step:QQbootstrap} Estimate uncertainty via bootstrap (\S\ref{sec:bootstrap}).
    \end{enumerate}
    \item\label{step:PP} For $d=1,\dots,D_\text{max}$, create a P--P plot:
    \begin{enumerate}
        \item Extract $\hat{\bm{y}}_d$ and $\bm{y}_d$ from $\hat{\bm{Y}}$ and $\bm{Y}$;
        \item For $q=1,\dots,Q$, calculate $\hat{y}_d^q$;
        \item Count $m_q$, the number of entries in $\bm{y}_d$ that are $<\hat{y}_d^q$;
        \item Plot $m_q/M$ vs.\ $q/(Q+1)$ for all $q$;
        \item\label{step:PPbootstrap} Estimate uncertainty via bootstrap (\S\ref{sec:bootstrap}).
    \end{enumerate}
    \item (Optionally) Visualise each principal eigenvector $\bm{v}_d$ as a bar plot (see example in Figure \ref{fig:pop_cosmos_broadband}).
    \item (Optionally) For $d=1,\dots,D_\text{max}$, compute the fraction of the total variance in $\bm{X}$ explained by its projection along the $d$th axis of $\hat{\bm{X}}$, $\var(\bm{y}_d)/\sum_{d=1}^D\var(\bm{y}_d)$.
    \item (Optionally) For $d=1,\dots,D_\mathrm{max}$, compute the two-sample K--S statistic between $\hat{\bm{y}}_d$ and $\bm{y}_d$ (\S\ref{sec:metrics}).
    \item\label{step:Wp} (Optionally) For $d=1,\dots,D_\mathrm{max}$, choose $p\geq 1$ and compute the $p$-Wasserstein distance $\mathcal{W}_p$ between the distributions of $\hat{\bm{y}}_d$ and $\bm{y}_d$ (\S\ref{sec:metrics}).
\end{enumerate}

\subsection{Selection and Calculation of Quantiles}
\label{sec:quantiles}
Here we elaborate further on the choice of $Q$ in step \ref{step:chooseQ} of \S\ref{sec:mathsteps}, and computation of quantiles in step \ref{step:quantiles}. We choose $Q=99$ -- i.e.\ a partitioning of the data into 100. In our examples (including the real galaxy study in \S\ref{sec:cosmos}), we have $N,M\approx100,000$ with $N\neq M$. It is a requirement that $Q\leq\min(N,M)$. For $N=M$, one can set $Q=N$ and trivially produce a Q--Q plot by plotting the sorted entries of $\bm{y}$ against the sorted entries of $\hat{\bm{y}}$. However, it is preferable to set $Q\ll\min(N,M)$. This makes generalisation to $N\neq M$ more straightforward. It also avoids excessive variance in the estimates of $\hat{y}^q$ and $y^q$ as $q\to 1$ and $q\to Q$. However, for estimating the $\mathcal{W}_p$ distance (step \ref{step:Wp} of \S\ref{sec:mathsteps}; \S\ref{sec:metrics}), it is preferable to use large $Q$.

The algorithm for computing $\hat{y}^q$ and $y^q$ is also a choice that must be made. For a comprehensive review of quantile estimation, see \citet{hyndman96}. For $\hat{\bm{y}}$ of length $N$, we can compute a pseudo-index $i=Nq/(Q+1)$ for the $q$th quantile $\hat{y}^q$. If $N/(Q+1)$ is an integer (i.e.\ if an exact partitioning of the dataset is possible), $\hat{y}^q$ will be the $i$th largest entry in $\hat{\bm{y}}$. Otherwise, an approximation must be used. The simplest and most intuitive method is a ``nearest rank'' approach where one calculates $g = i-\lfloor i\rfloor$ and takes the $\lceil i\rceil$th largest entry of $\hat{\bm{y}}$ if $g\geq0.5$, or the the $\lfloor i\rfloor$th largest entry if $g<0.5$. This produces a discontinuous estimator for $\hat{y}^q$. The default option in the \texttt{NumPy} library \citep{harris20}, which we use throughout this work, is a piecewise linear interpolation scheme proposed by \citet{gumbel39} and presented as definition 7 in \citet{hyndman96}. This produces an estimate of $\hat{y}^q$ that is continuous as a function of $q/(Q+1)$.

\subsection{Uncertainty Quantification}
\label{sec:bootstrap}
Here we outline algorithms based on bootstrapping \citep{efron79} that can be used to quantify uncertainty on Q--Q and P--P plots (steps \ref{step:QQbootstrap} and \ref{step:PPbootstrap} in \S\ref{sec:mathsteps}). A bootstrap estimator for the variance of sample quantiles was first introduced by \citet{maritz78}, with its properties proven by \citet{bickel81} and \citet{hall88}. For a data vector $\hat{\bm{y}}$ of length $N$, we can estimate the variance of the $q$th quantile $\hat{y}^q$ as follows:
\begin{enumerate}
    \item Select a number ($B$) of bootstrapped realisations to sample.
    \item For $b=1,\dots,B$:
    \begin{enumerate}
        \item construct a new vector $\hat{\bm{\gamma}}_b$ of length $N$ by sampling with replacement from $\hat{\bm{y}}$;
        \item estimate the $q$th quantile $\hat{\gamma}_b^q$ of $\hat{\bm{\gamma}}_b$ using the usual procedure (\S\ref{sec:quantiles}).
    \end{enumerate}
    \item Estimate the variance $\hat{\sigma}_q^2=(B-1)^{-1}\sum_{b=1}^B(\hat{\gamma}_b^q - \hat{y}^q)^2$.
\end{enumerate}
This allows us to show a confidence region on the diagonal of a Q--Q plot, thereby quantifying the significance of deviations. Similarly, for a P--P plot, we can estimate the variance of the cumulative probability estimates:
\begin{enumerate}
    \item Select a number ($B$) of bootstrapped realisations to sample.
    \item For $b=1,\dots,B$:
    \begin{enumerate}
        \item construct a new vector $\hat{\bm{\gamma}}_b$ of length $N$ by sampling with replacement from $\hat{\bm{y}}$;
        \item count $n_b^q$, the number of entries in $\hat{\bm{\gamma}}_b$ $<\hat{y}^q$;
        \item compute $p_b^q = n_b^q/N$.
    \end{enumerate}
    \item Find the variance $\hat{\sigma}_q^2=(B-1)^{-1}\sum_{b=1}^B[p_b^q - q/(Q+1)]^2$.
\end{enumerate}
In this way, it is also possible to show a confidence region on a P--P plot.

\subsection{Metrics}
\label{sec:metrics}
Here we include some details on computing the one-dimensional K--S statistic $\mathcal{D}_\mathrm{KS}$, and the $p$-Wasserstein distance $\mathcal{W}_p$. For two samples, $\hat{\bm{y}}$ and $\bm{y}$, of size $N$ and $M$ respectively, $\mathcal{D}_{\mathrm{KS}}$ is equal to the size of the largest difference between their empirical distribution functions $G_N(\hat{y})$ and $F_M(y)$. This can be computed as $\mathcal{D}_\mathrm{KS} = \sup_{y'}|G_N(y') - F_M(y')|$, with $G_N(y') = n_{<y'}/N$ and $F_M(y')=m_{<y'}/M$, where $n_{<y'}$ and $m_{<y'}$ are the number of entries in $\hat{\bm{y}}$ and $\bm{y}$ that are less than $y'$. Alternatively, a good approximation can be obtained quickly using the cumulative probabilities from \S\ref{sec:mathsteps} step \ref{step:PP}, $\mathcal{D}_\mathrm{KS} \approx \max_q|m_q/M - q/(Q+1)|$.

In one dimension, the $p$-Wasserstein distance between two distributions is given by $\mathcal{W}_p = \big\{\int_0^1 du'\,|G^{-1}(u') - F^{-1}(u')|^p\big\}^{1/p}$, where $G^{-1}(\hat{u})$ and $F^{-1}(u)$ are the inverses of the cumulative distribution functions $G(\hat{y})$ and $F(y)$ \citep[see e.g.][\S2.6]{peyre19}. We can estimate $\mathcal{W}_p$ from finite samples using any estimator for the inverse of the empirical distribution functions $G_N(\hat{y})$ and $F_M(y)$. For $N=M$, the computation is particularly straightforward \citep[e.g.][]{munk98, levina01}, with $\mathcal{W}_p\approx \big\{N^{-1}\sum_{n=1}^N|\hat{y}^{(n)} - y^{(n)}|^p\big\}^{1/p}$, where $\hat{y}^{(n)}$ and $y^{(n)}$ are the $n$th largest elements of $\hat{\bm{y}}$ and $\bm{y}$. For $N\neq M$, the same estimator can be used with a random subsample of length $\min(m,N)$ from the longer vector, or e.g.\ a Riemann sum over $Q$ quantiles $\mathcal{W}_p\approx \big\{Q^{-1}\sum_{q=1}^Q|\hat{y}^q - y^q|^p\big\}^{1/p}$. The quantiles estimated in \ref{sec:mathsteps} step \ref{step:QQ} can be recycled for this purpose, although one should check that $Q$ is large enough to yield a reliable estimate. We use the downsampling approach throughout this paper, but we have verified that the Riemann sum gives consistent results.
%
%
\bibliography{sample631}
\bibliographystyle{aasjournal}
%


\end{document}